\documentclass[twocolumn,number,5p]{elsarticle}

\usepackage{graphicx}
\usepackage{amsmath}
\usepackage{amssymb}

\newcommand{\be}{\begin{equation}} 
\newcommand{\ee}{\end{equation}}
\newcommand{\bea}{\begin{eqnarray}} 
\newcommand{\eea}{\end{eqnarray}}
\newcommand{\nin}{\noindent}

\begin{document}


\begin{frontmatter}
\title{IceRay: An IceCube-centered Radio-\v{C}erenkov GZK Neutrino Detector}

\author[uh]{P.~Allison}
\author[ohio]{J.~Beatty}
\author[taiwan]{P.~Chen}
\author[ucl]{A.~Connolly}
\author[uh]{M.~DuVernois}
\author[uh]{P.~Gorham}
\author[uw]{F.~Halzen}
\author[uw]{K.~Hanson}
\author[maryland]{K.~Hoffman}
\author[uw]{A.~Karle}
\author[uw]{J.~Kelley\corref{cor}}
\cortext[cor]{Presenter.}
\ead{jkelley@icecube.wisc.edu}
\author[uw]{H.~Landsman}
\author[uh]{J.~Learned}
\author[uh]{C.~Miki}
\author[uh]{R.~Morse}
\author[ucl]{R.~Nichol}
\author[ohio]{C.~Rott}
\author[uh]{L.~Ruckman}
\author[delaware]{D.~Seckel}
\author[uh]{G.~Varner}
\author[alabama]{D.~Williams}

\address[uh]{Dept.~of Physics and Astronomy, Univ.~of Hawaii, Manoa, HI
  96822, USA}
\address[ohio]{Dept.~of Physics, Ohio State University, Columbus, OH
  43210, USA}
\address[taiwan]{Dept.~of Physics, National Taiwan Univ., Taipei 106, Taiwan}
\address[ucl]{Dept.~of Physics \& Astronomy, Univ.~College London, London WC1E
  6BT, UK}
\address[uw]{Dept.~of Physics, Univ.~of Wisconsin, Madison, WI 53703, USA}
\address[maryland]{Dept.~of Physics, Univ.~of Maryland, College Park,
  MD 20742, USA}
\address[delaware]{Dept.~of Physics and Astronomy, Univ.~of Delaware,
  Newark, DE 19716, USA}
\address[alabama]{Dept.~of Physics and Astronomy, Univ.~of Alabama,
  Tuscaloosa, AL 35487, USA}

\begin{abstract}
We discuss design considerations and simulation results for IceRay, a
proposed large-scale ultra-high energy (UHE) neutrino detector at the South Pole.
The array is designed to detect the coherent Askaryan radio emission from UHE
neutrino interactions in the ice, with the goal of detecting the cosmogenic
neutrino flux with reasonable event rates.  Operating in coincidence with
the IceCube neutrino detector would allow complete calorimetry of a subset of the
events.  We also report on the status of a testbed IceRay station which
incorporates both ANITA and IceCube technology and will provide year-round
monitoring of the radio environment at the South Pole.  
\end{abstract}


\end{frontmatter}


\section{\label{sec:intro}Introduction}

\begin{figure}
\includegraphics[width=0.9\columnwidth]{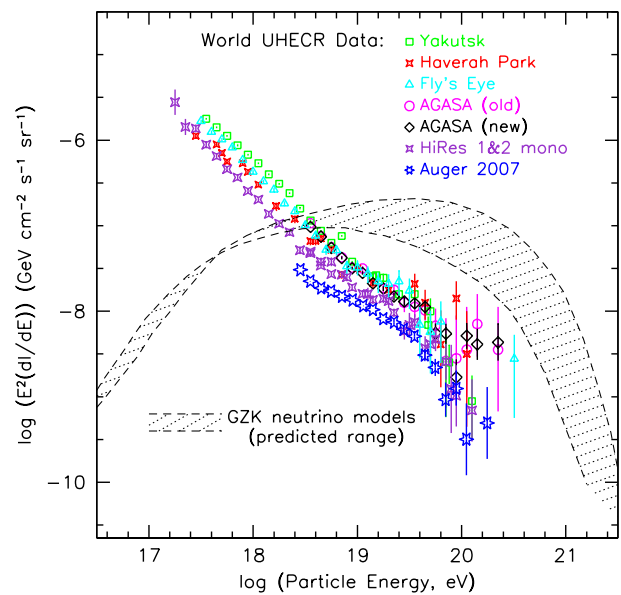}
\caption{\label{fig:uhecr}
World ultra-high energy cosmic ray and
predicted cosmogenic neutrino spectrum as of early 2007,
including data from the Yakutsk~\cite{Yakutsk},
Haverah Park~\cite{Haverah}, the Fly's Eye~\cite{Flyseye},
AGASA~\cite{AGASA}, HiRes~\cite{Hires07}, and Auger~\cite{Auger07},
collaborations. Data points represent differential
flux $dI(E)/dE$, multiplied by $E^{2}$. Error bars are statistical only.
GZK neutrino models are from Protheroe \& Johnson~\cite{Proth_gzk} and
Kalashev {\em et al.}~\cite{Kal02}.}  
\end{figure}

Continued progress in the determination of the ultra-high energy cosmic ray
(UHECR) spectrum above $10^{17}$ eV has established the presence of the
Greisen-Zatsepin-Kuzmin (GZK) suppression \cite{Hires07}, resulting from the interaction of UHECRs
with the cosmic microwave background (see Fig.~\ref{fig:uhecr}).  Such
interactions lead to a ``guaranteed'' flux of UHE neutrinos, although the
characteristics of the flux depend on the details of the source distribution, UHECR composition,
and other currently unknown factors.  Measurement of the GZK neutrino flux
would not only shed light on these issues, but also could indicate the
UHECR sources themselves via the direction of the individual neutrinos.

An array to detect UHE neutrinos via their coherent radio emission in a dense
medium was originally described by Gusev and Zhelezhykh \cite{gusev},
based on theoretical work by Askaryan.  Since then, significant experimental
work by the RICE collaboration \cite{RICE03} has established many of the
fundamental characteristics of radio transmission in the polar ice, and the
ANITA balloon experiment \cite{anitalite} has currently set the best
limits on UHE neutrino fluxes.  Direct observation of coherent radio
emission using an ice target at SLAC has also confirmed the theoretical
foundations described by Askaryan \cite{slac_ice}.

The expected flux of GZK neutrinos is nevertheless quite small, requiring a
large array (or in the case of ANITA, a huge target) to ensure reasonable
event rates.  We propose to extend the IceCube neutrino detector \cite{icecube} to energies from $10^{17}$ to
$10^{20}$ eV with a sparse array of radio antennas (IceRay).  An initial array of 50
$\text{km}^2$ is designed to provide event rates of $O(\text{few/yr})$ and
establish the baseline flux level, while a final target array of 300 to
1000 $\text{km}^2$ could provide $O(100)$ events per year.  Centering the
array around IceCube allows a subset of the events to be detected in
coincidence, providing complete calorimetry of both the initial interaction
(via the radio emission) and the outgoing lepton (via the optical
emission).  While rare, such events provide a valuable means of
cross-calibration and reduction of systematics in the absolute energy
scale. 


\section{\label{sec:design}Design Considerations}

We discuss several of the design considerations for a large-scale radio
array in the ice, in particular the operating frequency and geometry.  

\subsection{Operating Frequency}

We are initially bounded in operating frequency to the region between
several MHz, where backgrounds may be prohibitively large, to around 1 GHz,
where the ice becomes opaque.  The coherent emission from an Askaryan pulse
has a peak field strength which rises linearly with frequency, but the received voltage
at an antenna is inversely proportional to frequency, so the direct
dependence on frequency cancels when considering the signal-to-noise
ratio (SNR).  However, a dependence on the bandwidth remains; specifically, we
find 

\be
\text{SNR} \propto E_{\text{shower}} \sqrt{\frac{G\Delta f}{kT_{\text{sys}}Z_0}}\ 
\ee

\nin for a shower of energy $E_{\text{shower}}$, using a receiver with
gain $G$ and noise temperature $T_{\text{sys}}$, and where $Z_0$ is a reference
impedance.  Therefore, high bandwidth is important, but there is no direct
dependence on the center frequency of the band.

Other considerations, however, indicate a preference for lower
frequencies.  First, while the peak field strength of the \v{C}erenkov
emission rises with frequency, the angular width of the \v{C}erenkov cone
gets narrower \cite{lehtinen}.  Effectively, this reduces the total solid
angle available for detection at high frequencies.  

Furthermore, the frequency dependence of the attenuation length of the ice
itself plays an important role.  Over the 200-700 MHz range, the
attenuation length decreases by approximately 25-30\% \cite{barwick_ice}.  Because the
effective volume, to first order, varies as $L_{\text{atten}}^3$, this
implies a strong loss at high frequencies.  The overall conclusion is that
a high bandwidth, low frequency approach is optimal.  Given that a bandwidth
factor of 5 is reasonably achievable, we set a preliminary target frequency
range of 60-300 MHz.  

\subsection{Geometry}

\begin{figure}[htb!]
\begin{center}
\centerline{\includegraphics[width=3.4in]{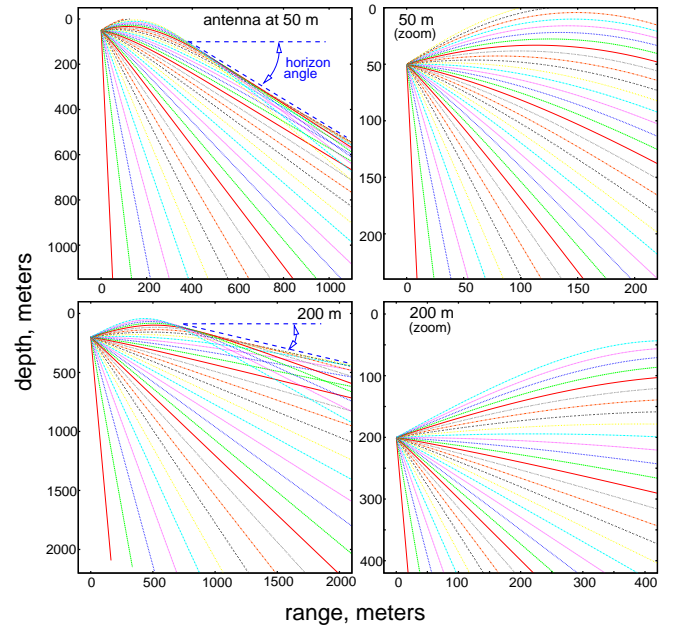} }
\caption{Example of refraction effects for shallower antenna locations.
Both 50~m (upper) and 200~m (lower) deep antenna locations are shown.
On the left are the wide-scale ray geometries, showing the terminal
horizon angle in each case, and on the right the details of the
ray bending in the near zone are shown. \label{fig:raytrace}}
\end{center}
\end{figure}

Because the radio field attenuation length in ice is of $O(1\ \text{km})$
\cite{barwick_ice}, one can cover a relatively large area somewhat
sparsely.  While deploying detectors on the surface is the most
cost effective, refraction effects greatly penalize the volumetric
acceptance.  The index of refraction varies from 1.79 in the deep
ice (below about 200m) to 1.33 in the packed snow at
the surface \cite{RICE03}.  The low-density region is known as the
\textit{firn}, and upward-going rays moving through this region are bent
away from the surface.  This creates a horizon angle --- that is, shallower rays cannot
reach the detector.  This angle gets much less severe as one moves deeper
into the ice (see Fig.~\ref{fig:raytrace}), suggesting that deploying
antennas in holes, say, 50m or 200m below the surface is much more
efficient.   

IceCube has already developed drilling technology that can be utilized 
for IceRay.  While the enhanced hot water drill (EHWD) used for
drilling the 2.5 km deep holes for IceCube string deployment is not mobile
enough for our purposes, the independent firn drill (IFD) which drills the
``pilot'' holes for the EHWD can be easily moved.  The IFD is a
``hotpoint''-style drill which melts into the firn using a cone of
closed-loop 
copper tubing, heated with a propylene glycol/water mixture.  The IFD
currently can drill at about 4 m/hour, with an average power usage of
approximately 100 kW.  The IFD is effective to depths of 40-50 m,
after which pooling water causes power 
usage to spike.  Adding a pump to extract this water is a simple
modification which could alleviate this issue.  Ultimately, we expect that
drilling to 200 m is logistically manageable and cost effective, either
with a modified IFD or other technology.

\subsection{Baseline Configurations}

\begin{figure*}[t]
\begin{center}
\centerline{\includegraphics[width=2.8in]{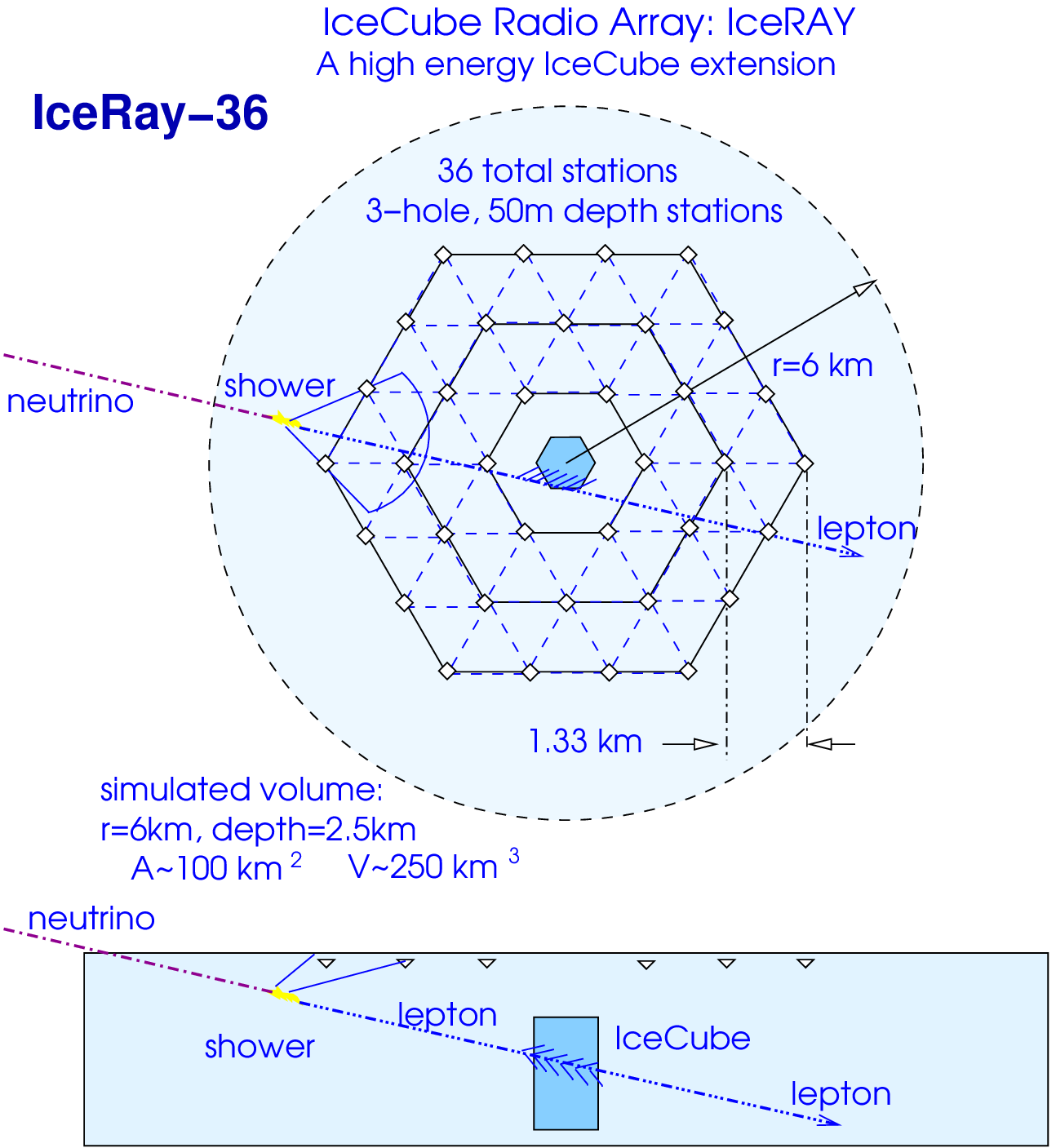}~~\includegraphics[width=2.8in]{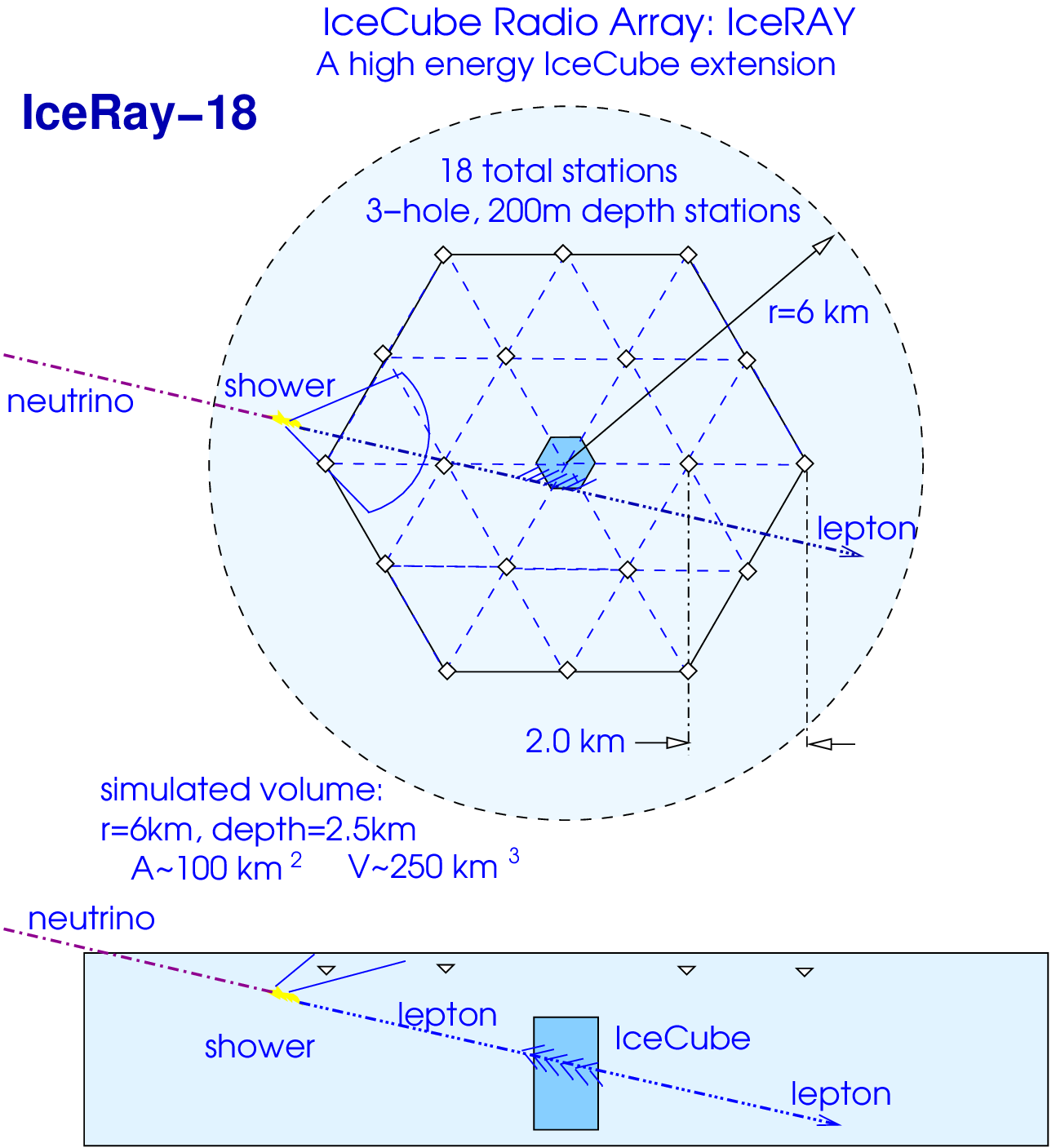} }
\caption{Left: Baseline 36-station, 50 m depth array, in a plan view (top) and side view
(bottom) showing the simulated interaction region around the detector.
Right: Alternative 200 m depth, 18-station array. \label{fig:geom}}
\end{center}
\end{figure*}

Given the above design considerations, we focus on two geometries for
the initial 50 $\text{km}^2$ phase of the array: a shallower, denser array
deployed at a depth of 50 m and with 36 stations; and a deeper, more sparse
array deployed at a depth of 200m and with 18 stations.  The configurations
are chosen to have approximately the same cost and volumetric acceptance in
the peak energy region of the GZK neutrino flux, around $10^{18}$ eV.
Figure \ref{fig:geom} shows the station arrangement in more detail.  Each
station consists of three holes separated by 5-10 m, with four antennas
(two of each polarization, horizontal and vertical) in each hole.
Directionality is achieved for even single-station events via timing
information from these local baselines.   

\section{\label{sec:sim}Simulated Event Rates}

\begin{figure}
\includegraphics[width=0.9\columnwidth]{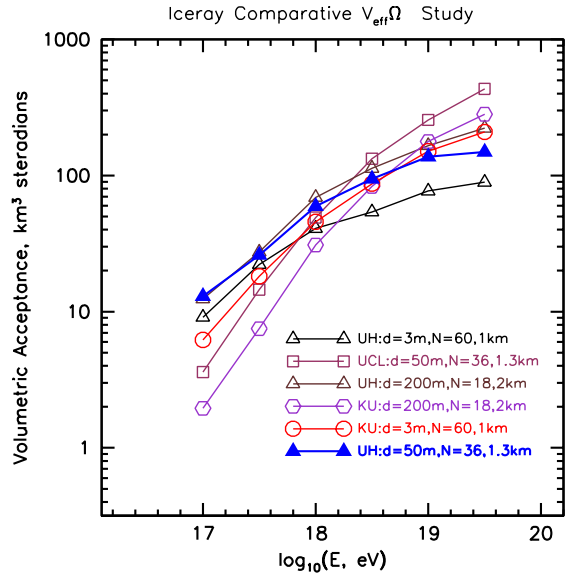}
\caption{\label{fig:veff} Volumetric acceptance, in km$^3$ steradian, of
various array configurations ($d$ = station depth; $N$ = number of
stations; and station spacing), including results from three
independent simulation chains.}
\end{figure}

The primary IceRay simulation chain is based on Monte Carlo code developed
for ANITA and SalSA \cite{salsa}, but independent crosschecks have been performed with
ARIANNA \cite{arianna} and RICE simulation chains.  The volumetric acceptance of different
array configurations is shown in Fig.~\ref{fig:veff}, and we note
reasonable agreement in the important energy range of $10^{18}$ eV.  In
general, the 18-station deep configuration gives higher acceptance than the
36-station shallow configuration at the higher energies, but drops off
at low energies due to the increased station spacing.  

Table \ref{table:rates} shows integrated event rates for the two
baselines configuration studied.  ``Standard'' fluxes, such as ESS with a
$\Lambda$CDM cosmology \cite{ess} and Protheroe \cite{Proth_GZK}, result in
approximately 3-10 events 
per year, but as mentioned earlier, this could vary almost an order of
magnitude in either direction depending upon UHECR composition and source
evolution.  Iron UHECR models in particular tend to produce significantly lower rates
\cite{Ave05}, although these are currently disfavored by measurements of
the spectral endpoint \cite{Hires07}.  An important point is that no
irreducible backgrounds are expected, so detection of even a few events
would be significant.   

\begin{table}[hbt!]
\caption{\label{table:rates}
Event rates per year for several classes of UHE cosmogenic
neutrino models.  The ``36-50'' rates are for the 36-station, 50m-deep
configuration, and the ``18-200'' rates are for the 18-station, 200m-deep
configuration.} 
\begin{center}
\begin{small}
  \begin{tabular}{|c|c|c|} \hline
    Cosmogenic neutrino model &36-50 &18-200 \\
    & ev/yr & ev/yr \\ \hline  \hline
    Fe UHECR, std. evolution & 0.50 & 0.60 \\ \hline
    Fe UHECR strong src. evol. & 1.6 & 1.8 \\ \hline
    ESS 2001,$\Omega_m = 0.3$, $\Omega_{\Lambda} = 0.7$ & 3.5  & 4.4    \\\hline
    Waxman-Bahcall-based GZK-$\nu$ flux   &4.2  & 4.8  \\\hline
    Protheroe and other std. models   &4.2-7.8  & 5.5-9.1  \\\hline
    Strong-source evolution (ESS,others)   & 12-21  & 13.8-28   \\ \hline
    Maximal, saturate all bounds& 24-40  & 32-47  \\ \hline
  \end{tabular}
\end{small}
\end{center}
\end{table}

\begin{figure}[htb!]
\begin{center}
\centerline{\includegraphics[width=0.9\columnwidth]{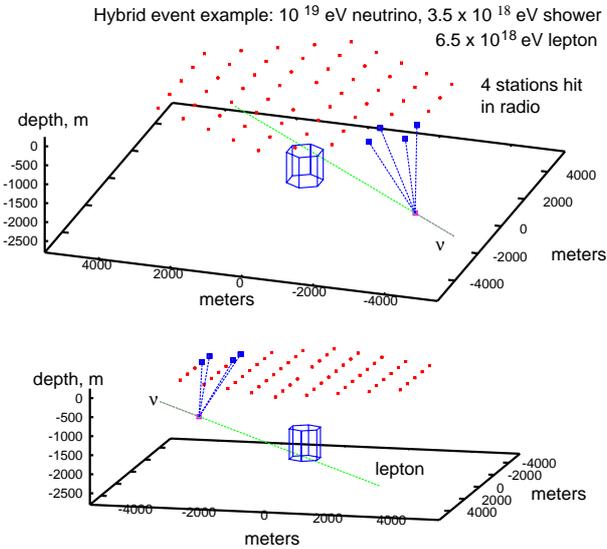} }
\caption{Example of a hybrid event where the vertex is seen by
four surface radio detectors and the resulting
lepton passes near enough to IceCube to make a detection. \label{fig:hybrid}}
\end{center}
\end{figure}

One significant motivation to build IceRay at the South Pole is to allow
for the possibility of coincident, or ``hybrid'' events with the IceCube
detector.  A $\nu_\mu$ or $\nu_\tau$ event can produce both an initial shower
and a long-ranged charged lepton with the potential for
detection in both radio and optical channels.  A typical geometry for such
a hybrid event is shown in Fig.~\ref{fig:hybrid}. 

Such a hybrid event allows cross-calibration of the energy scale of either
detector, and while such events are rare, they are background-free.  Event
rates per ten years for various GZK flux models are shown in table
\ref{table:hybrid}.  Adding a high energy ``guard ring'' of strings to
IceCube (the ``IceCube-plus'' configuration; see Ref.~\cite{IceCubeplus})
increases the hybrid event rate by up to a factor of two.  We have
also conservatively assumed here that each detector triggers independently;
adding sub-threshold cross-triggering would also increase the event rate.  

\begin{table}[hbt!]
\caption{\label{table:hybrid} Hybrid event rates for the baseline IceCube, and
IceCube-plus (1.5~km guard ring), per 10 years of operation,
for several classes of UHE cosmogenic
neutrino models, assuming the IceRay-36, 50m-deep radio array.}
\begin{center}
\begin{small}
  \begin{tabular}{|c|c|c|} \hline
    Cosmogenic neutrino model & IceCube & IceCube+   \\
    & 10 yrs & 10 yrs \\ \hline  \hline
    ESS 2001$\Omega_m = 0.3$, $\Omega_{\Lambda} = 0.7$ & 3.2  & 6.4    \\\hline
    Waxman-Bahcall-based GZK-$\nu$ flux   & 3.8  & 7.6 \\\hline
    Protheroe and other standard models   &3.8-7.1  & 5.0-8.2  \\\hline
    Strong-source evolution (ESS,others)   & 10-19  & 13-25   \\ \hline
    Maximal fluxes, saturate all bounds& 22-36  & 30-44  \\ \hline
  \end{tabular}
\end{small}
\end{center}
\end{table}


\section{\label{sec:testbed}Testbed Station}

Recent data from the ANITA flights have demonstrated that the South Pole is
not a particularly radio-quiet environment (at least in the austral
summer), but no capability currently exists for year-round monitoring 
of the natural and anthropogenic backgrounds.  To understand and
characterize these backgrounds, as well as to test prototype hardware, we have built the
IceRay testbed station.  

\begin{figure}
\includegraphics[width=0.9\columnwidth]{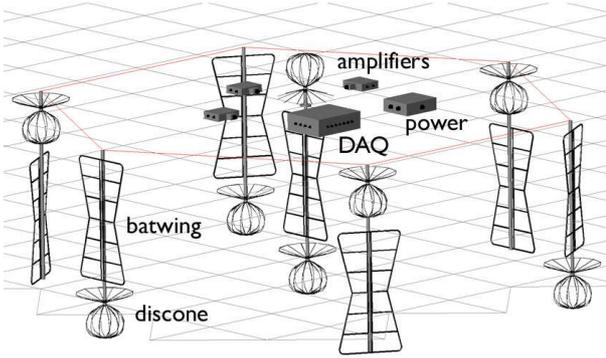}
\caption{\label{fig:iceray_cluster}Layout of the IceRay testbed station.}
\end{figure}

When deployed, the testbed will be a single surface station with pairs of antennas buried in
shallow boreholes (2.5 m deep) in the snow.  The boreholes are arranged in
a circle of radius 5 m.  Each hole contains a \textit{discone} antenna optimized for
vertically polarized signals, and a \textit{batwing} antenna for horizontally polarized
signals. Figure \ref{fig:iceray_cluster} shows the layout of the station as
well as the antenna geometry.

\begin{figure}
\includegraphics[width=0.9\columnwidth]{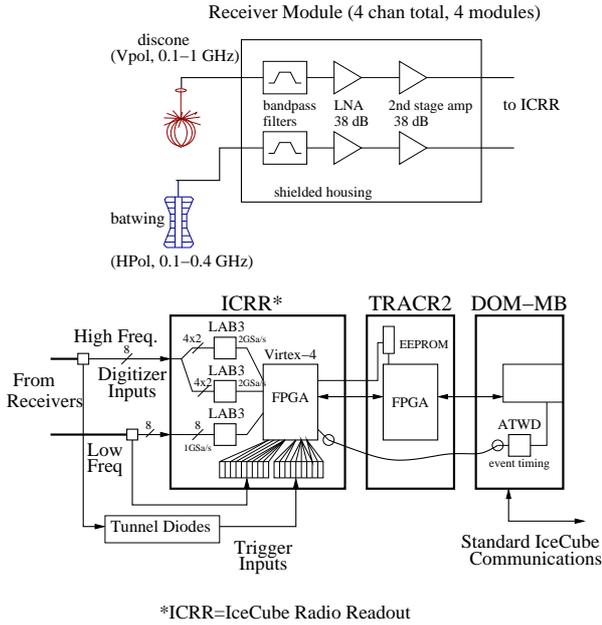}
\caption{\label{fig:testbed_daq}Schematic representation of the data
  acquisition system for the IceRay testbed station.}  
\end{figure}

The data acquisition system (DAQ) combines hardware elements of ANITA,
IceCube, and the digital radio modules of AURA (Askaryan Under-ice Radio
Array; see Ref.~\cite{aura}).  Four antennas  (two of each polarization)
are first fed into a low-noise amplifier chain 
(with a total gain of $\sim76$ dB) in a shielded housing.   The combined system has a
bandpass of 115 MHz-1.2 GHz.  
High- and low-frequency components are split and separately digitized with
LABRADOR3 ASICs \cite{Varner4} at 2 GSa/s and 1 GSa/s, respectively, as
part of the IceCube Radio Readout board (ICRR).  The digitized waveforms are buffered and transferred through an
intermediary board, the TRACR, to a standard IceCube digital optical module
mainboard (DOM-MB), which also provides event
time-stamping via its own digitizer, the ATWD.  The DOM-MB communicates via
the standard IceCube communications protocol, so the station can be
connected to the IceCube cabling network and controlled from the IceCube
counting house.  Furthermore, standard IceCube time calibration procedures
can synchronize the DAQ to a GPS clock.  Figure \ref{fig:testbed_daq} shows
a schematic of the DAQ components. 

\section{\label{sec:outlook}Outlook}

We continue to study open issues with the design of the full array (of
order 1000 $\text{km}^2$), such as power distribution and communications.
We are also actively pursuing efforts to increase the sensitivity of the
system down to the cosmogenic $kT$ noise floor of about -114 dBm/MHz, as well
as working with South Pole station management to control the anthropogenic
noise in the 60-1000 MHz frequency range, in order to lower our energy
threshold below $10^{17}$ eV.  This would both increase the total event
rate and provide enhanced opportunities for hybrid events.  Furthermore,
other techniques such as acoustic detection of UHE neutrinos are developing
rapidly, suggesting that a hybrid radio-optical-acoustic array may have
significant benefits for systematics and cost \cite{hybrid}.

Installation of the IceRay testbed in the austral summer of 2009
will allow precise characterization of the noise environment and will
facilitate further development of the 50 $\text{km}^2$ array.  IceCube
construction will complete in 2011, and we hope to phase in
construction of IceRay at that time, as the ability to use IceCube as the
core of a GZK neutrino detector is an unparalleled opportunity.


\end{document}